\documentclass{emulateapj}
\usepackage{natbib}

\shorttitle{Cloudshine}
\shortauthors{Foster and Goodman}

\begin{document}


\title{Cloudshine: New Light on Dark Clouds}


\author{Jonathan B. Foster and Alyssa A. Goodman}
\affil{Harvard-Smithsonian Center for Astrophysics, 60 Garden Street, Cambridge, MA 02138}


\begin{abstract}
We present new deep near-infrared images of dark clouds in the Perseus molecular complex. These images show beautiful extended emission which we model as scattered ambient starlight and name ``cloudshine''. The brightness and color variation of cloudshine complicates the production of extinction maps, the best tracer of column density in clouds. However, since the profile of reflected light is essentially a function of mass distribution, cloudshine provides a new way to study the structure of dark clouds. Previous work has used optical scattered light to study the density profile of tenuous clouds; extending this technique into the infrared provides a high-resolution view into the interiors of very dense clouds, bypassing the complexities of using thermal dust emission, which is biased by grain temperature, or molecular tracers, which have complicated depletion patterns. As new wide-field infrared cameras are used to study star-forming regions at greater depth, cloudshine will be widely observed and should be seen as a new high-resolution tool, rather than an inconvenience.

\end{abstract}


\section{Introduction}

The mass structure of molecular clouds can be studied in a variety of ways, but none is without problem. Molecular line maps can produce high-resolution views, but depletion and excitation variations confound conversion to true column density. Dust emission at sub-millimeter wavelengths depends on the emitting characteristics of grains, which may vary across molecular clouds. The most direct method for determining the mass structure is extinction mapping, exemplified by NICER (Near-Infrared Color Excess method Revisited) (\citet{Lombardi2001}) which uses the reddening of background stars in the near-infrared to infer the dust column density. The resolution of this method is proportional to the number of background stars visible through the cloud: this is a function of galactic position, the quantity of obscuring material, and the depth of the images used.

In this letter we report on near-IR observations originally made to construct an extinction map in the outer regions of the Perseus molecular cloud at moderate column densities (A$_V$ $<$ 30 mag) and relatively high galactic latitude. In these deep images (complete to J = 20.1, H = 20, Ks= 19.3), there is pronounced emission structure associated with the clouds (see Figures 1 and 2). This diffuse structure, or ``cloudshine'', is well explained by scattering of the diffuse interstellar radiation field by dust grains with the optical properties of \citet{Weingartner2001}. With deep enough images, it seems likely that every dark cloud becomes a faint reflection nebula in the near-IR, a phenomenon which has been reported a few times before (\citet{Lehtinen1996} and \citet{Nakajima2003}) in the context of constraining dust grain models. The structure's ubiquity, relative brightness, and color variation on arcsecond scales complicate the production of extinction maps from deeper exposures where the emission from the extended structure is roughly equal to that of faint background stars. However, the cloudshine, while less easy to interpret than extinction, presents a new way to study the mass structure of molecular clouds at sub-arcsecond resolution.

\section{Observations}

The observations were taken in January 2005 at Calar Alto Observatory on the 3.5 meter, using the new OMEGA 2000 camera. This wide-field near infrared camera has a 2048x2048 pixel array, and provides a 15.1x15.1\arcmin  field of view. We covered 6 science fields and 2 control fields in three nights of good weather with $\sim$ 0.6\arcsec  seeing. We observed with long exposure times at H (45 minutes) and Ks (37 minutes) in order to produce the extinction map, and added short exposures at J (6 minutes) to check the reddening law, although these proved useful for studying the diffuse emission. We covered the entirety of B5 with a mosaic of four frames (J-band images for the central two frames were lost due to a pointing error with the telescope); two dark clouds to the south-east of Perseus, L1448 and L1451; and two control fields outside the known boundaries of the cloud. All science fields showed cloudshine, while neither control field did. The control fields were observed with the same exposure times as the science frames, and reached the same depths. Additional images and data not in this letter are available through the COMPLETE (CoOrdinated Molecular Probe Line Extinction and Thermal Emission) Survey website at http://cfa-www.harvard.edu/COMPLETE .

We did not anticipate significant extended emission, and thus observed without taking off-target sky frames for each exposure. Instead, we dithered (using a semi-random pattern) around a pointing, obtaining exposures of two or three seconds each. After dark subtraction and division by a flat-field, we used the median images from the 3 images taken before and after a frame to construct a sky. The implementation of this was carried out by XDIMSUM under IRAF, although the same process was run through ESO's ECLIPSE software to check for consistency. As the extent of the cloudshine was larger than the dither pattern ($\sim$ 30\arcsec), our images suffer from self-sky subtraction. We studied the effect of this by creating synthetic data with known shapes, adding on a sky frame derived from data with some random noise, and running this sythetic data through the same sky subtraction routine in XDIMSUM. In general, the effect of self-sky in such a well-dithered pattern was a small ( $<$10\%) reduction in surface brightness in a band about the size of the dither pattern around edges of a feature and a dark rim around the edge of structure. We thus assured ourselves that self-sky did not substantially alter the morphology of our extended structure. However, future observations to study this phenomenon should take off-target frames to obtain more accurate sky-subtraction. 

Because these observations were taken as part of the COMPLETE Survey (\citet{Goodman2004}; \citet{Ridge2005}), we had immediate access to a variety of other images of the same region. The diffuse emission traced well the location of the 850 micron and 1.2-mm dust continuum emission, and N$_2$H$^+$ and $^{13}$CO line emission, so it was clear that cloudshine was associated with the dense portions of the clouds. We made color images to aid in studying the data. 2MASS stars within the field were used to convert flux/pixel to MJy/sr, and the component images were weighted so that a pixel's color shows the ratio of the fluxes within each band coming from that region of sky. In Figure 1, we have shown L1448 with each component image scaled to the same flux scale in MJy/sr, and the region nicely highlights several different physical mechanisms. The outflows are red (Ks), the reflection nebula is blue (J) and the cloudshine is brownish (predominately H and Ks). In Figure 2 we show L1451 with the same weighting of the color components by flux, but different optimization of contrast and black-point. Additionally, we have overlaid 1.2-mm continuum dust emission from COMPLETE (\citet{Tafalla2006}) to show how the reddest sections of the cloudshine correspond to the regions of high thermal dust emission, and how the edges of the cloudshine trace the contours of the 1.2 mm emission. 

\section{Discussion}

\subsection{Explanations considered and rejected} 

We considered a variety of possible sources for the cloudshine. L1448 shows a number of bright outflows in the Ks image associated with shocked H$_2$ (v=1-0 transition); could weaker turbulent shocks excite H$_2$ levels in a similar way? \citet{Black1987} present a full spectrum of H$_2$ lines in their Figure 2. The vast majority of strong lines arise only from UV excitation, which is unlikely in dark clouds. Furthermore, the integrated intensity of these lines is significantly greater in J than in H, while our cloudshine is much stronger in H than J (more than can be explained by a few magnitudes of visual extinction between us and the source). Even more unlikely would be thermal emission from the grains. To radiate most strongly between H and K requires 1500 K, or shocks traveling on the order of 10 km/s. Stochastic heating of grains \citep{Sellgren1984} by individual UV photons cannot be ruled out, but the preferential emission from high column density regions indicates that this is certainly not the main mechanism.

New emission features of dust are a possibility. \citet{Gordon2000} report on a strong feature at 1.6 microns (H-band) seen in spectra around the reflection nebula NGC 7023. Without spectra of the cloudshine it is impossible to rule out that such a feature might be contributing to the color of the dust, but this feature is tentatively identified as photoluminesence from amorphous iron disilicide ($\beta$-FeSi$_2$), so again, the low UV flux in the portions of these dark clouds which give rise to cloudshine makes this effect unlikely to contribute strongly.

\subsection{Modeling the diffuse structure as reflected starlight}

The simplest explanation is reflected star-light. The appearance of clouds in reflected starlight is a function of the incident radiation field, the optical properties of the grains, and the mass structure of the cloud. This problem becomes simplified in dark clouds far from stellar clusters since the radiation field is approximately uniform locally and is a simple function of galactic latitude. Also, it is difficult to strongly constrain grain models in the near-IR -- a relatively wide range of parameters produces roughly the same appearance (\citet{Lehtinen1996}, \citet{Nakajima2003}). Conversely, the exact properties of the dust have little influence on the appearance of the scattered light, leaving us with an image that depends mostly on the mass structure of the cloud. 

To test the reflection hypothesis, we used a Monte Carlo approach to model the appearance of spherical clouds in the near infrared. Our study closely followed the work of \citet{Mattila1970} and \citet{Witt1974} who modeled arbitrary density profiles and optical properties of the dust. In our code, much like in \citet{Witt1974}, a sphere is uniformly illuminated along one axis by a number of photons. These photons enter a cloud consisting of a series of shells of different density, and hence optical depth. The length before scattering is chosen randomly to produce the correct distribution, and if this length exceeds the distance along the current trajectory to the next shell, the photons passes into this shell, adjusts the remaining optical path length for the new density environment, and continues. When the photon scatters, its weight is reduced by the albedo for the grains, and a new direction is chosen from the Henyey-Greenstein distribution with a factor g = $<\cos{\theta}>$ characterizing the amount of forward scattering. The Henyey-Greenstein function is inaccurate at wavelengths far from the optical, where \citet{Draine2003} advocates using a new function, but for the near-IR the error in the HG function is less than 2\%.

Photons eventually escape from the cloud, and their final direction cosines with respect to their initial approach are binned into equal area bins. The problem is then inverted. Since we track each photon along its life-history, we weight each entering photon by the local weight of the photons in its final bin.  We re-weight photons to make each solid angle bin around the sphere have the same weight of photons leaving through it (in the physical problem this corresponds to isotropic illumination), and we then bin the incoming photons by initial impact parameter, giving us a radial profile of the cloud.

By choosing the correct weighting for each individual density shell, we can set up an arbitrary density profile and total optical depth for the cloud. Choosing a certain depth in visual extinction, and adopting the typical ratio of reddening to visual extinction for dark clouds (R$_{V}$ = 5.3 \citet{Weingartner2001}), we obtain total optical depths to the center of the cloud for each of J, H, and Ks. We run the simulation three times, corresponding to the same physical cloud and dust grain model (but different optical depths, albedos, and forward scattering values corresponding to the different wavelengths) and thus build a radial profile for comparison. We must then include the transmitted light which passes through unscattered. This can be calculated analytically for a specified density distribution. This is added to the scattered light profile, and then the illuminating source of light is subtracted out to mimic the effect of sky subtraction.

The final input necessary is the ambient starlight, and here only a rough guess is possible. \citet{Lehtinen1996} model the diffuse galactic light as a function of galactic latitude based on J and K data from COBE (COsmic Background Explorer) on DIRBE (Diffuse InfraRed Background Experiment). They fix the brightness at H so that the colors (H-K and J-H) are the same as the mean color of other spiral galaxies. We adopt their mean surface brightnesses of $I_J = 0.44$, $I_H=0.61$, and $I_K=0.48$ MJy/sr. Since our core lies near the center of a larger cloud surrounding dust (most clearly seen in the cloudshine itself in Figure 2), we reddened this galactic flux by 5 A$_{V}$, the approximate column density of the area as known from 2MASS extinction maps (\citet{Ridge2005}).

As shown in Figure 3, we selected one roughly circular dark core in L1451 to test this model, and made a radial plot of the brightness level above the sky. We modeled the scattered light resulting from power law density profiles ranging from r$^{-1}$ to r$^{-3}$. The steeper the density profile, the narrower the peak produced in the model, as most radiation is scattered from a small range of optical depth around one. Although not an exact fit, an r$^{-2}$ density profile and a total optical depth of 120 magnitudes of visual extinction broadly reproduces the shape and magnitudes of the three radial profiles, but is a less good fit in the inner portions. One possibility is that the linearly spaced shells do an increasingly poor job of tracing the density profile at the center; though the physical system must also have a flattened density profile at the center. An exact fit is probably impossible because the core is not truly spherical, is probably not isotropically illuminated and self-sky subtraction will somewhat alter the radial profile.

This type of modeling was introduced by \citet{Witt1990} who used reflected optical light and an extinction profile to study the structure of dark clouds. Extending this to the infrared has two main advantages. First, deep near-infrared images of dark clouds are likely to be obtained for a number of other reasons (e.g. extinction mapping or studying the embedded population). Second, near-infrared radiation penetrates much deeper into clouds, giving a more complete image of their structure.

\subsection{Studying the mass structure of dark clouds}

Only in simple cases such as described in the previous section will it be possible to infer the 3D structure of a cloud from cloudshine, but an exciting new work by \citet{Padoan2005} shows that scattered light images may lead directly to a 2D column density map. By illuminating a model cloud, they demonstrate a method by which the color of scattered light be be directly used as an unbiased estimator of the column density. This method raises the possibility of subarcsecond column density maps, a typical linear resolution improvement (in these images) or 100 times over extinction mapping and 20 times over molecular line or dust emission mapping. 

A final application of cloudshine is in comparing models to data. Simulations of turbulent structure produce mass density models of clouds, which can then be illuminated with IR photons and the resulting image can be statistically compared with our images in a variety of ways (e.g. Principle Component Analysis or the Spectral Correlation Function), to test the validity of the simulations.

\section {Conclusion}
As near-infrared cameras and space-born instruments (such as the James Webb Space Telescope) become more efficient, cloudshine will be widely observed in deep exposures. Although it complicates observing strategy from the gound, and makes accurate color photometry of faint sources more difficult, cloudshine is also a useful tool. We have presented here one example of how we may study the density structure of dark clouds where other tools cannot, and we have pointed out how wide-scale near-infrared images of this phenomena could lead to constraints on models of turbulence. Within the fairly large assumptions required, cloudshine can be adequately explained purely as scattered light, although other contributions cannot be firmly ruled out. Isolated, spherical clouds would be the best place to test these models in detail, while larger, more complex clouds provide more interesting comparisons with theoretical models.

\acknowledgments

We would like to thank Elisabeth Adams and Yuri Beletsky for assistance during observing, and Ed Churchwell, Bruce Draine, Felice Frankel, Tracy Huard, Phil Myers, Paolo Padoan, Alicia Porras, Ellen Zweibel, and the full COMPLETE team (especially Jo\~ao Alves) for comments on this work while it was in progress.  We particularly thank Paolo Padoan for providing an advance copy of his numerically-oriented companion article to this one. This work was supported by grants from both NASA and the NSF to JBF and AAG. 




\bibliographystyle{apj}

\clearpage

\begin{figure}
\figurenum{1}
\plotone{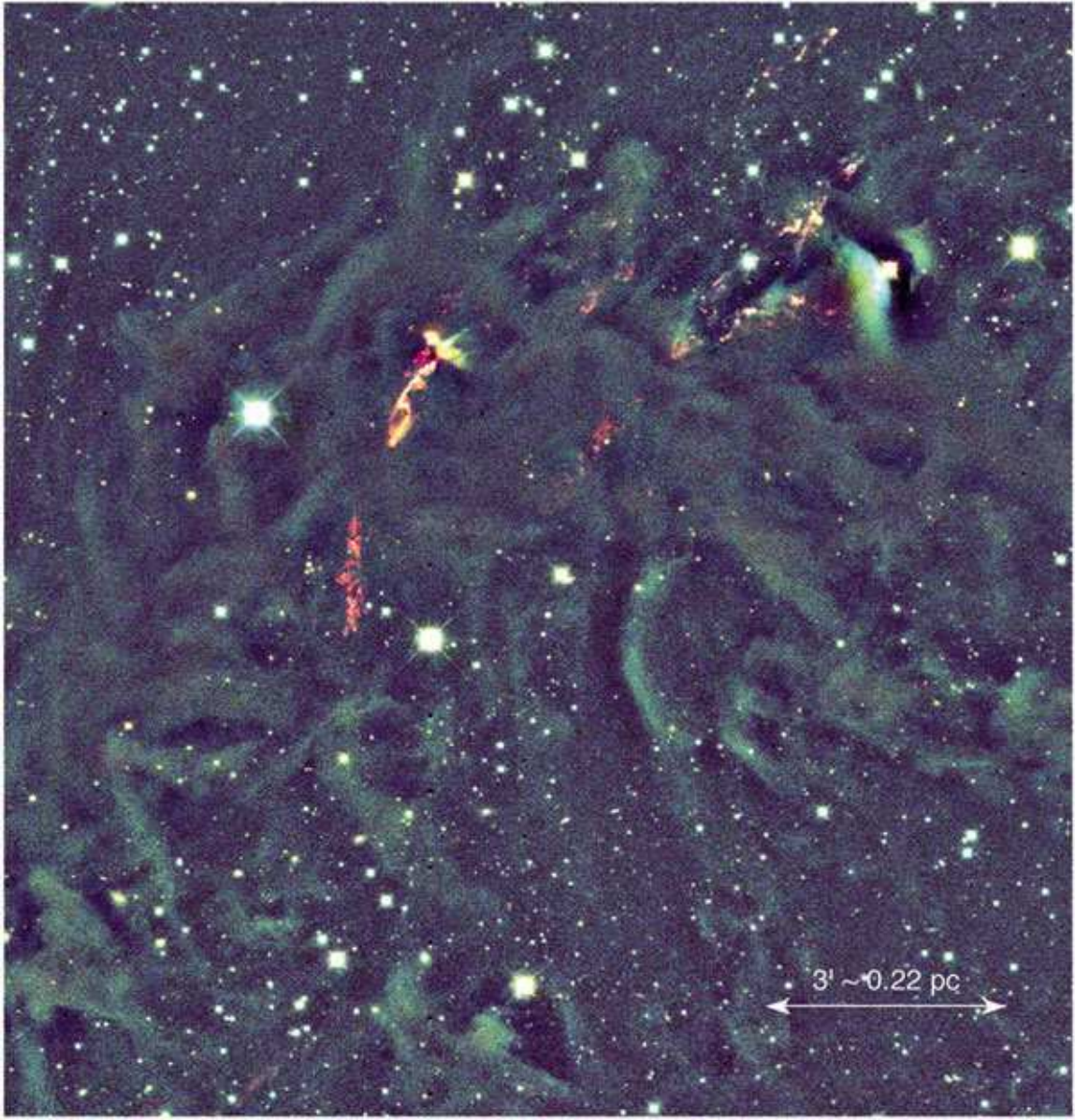}
\caption{L1448 in false color. Component images have been weighted according to their flux in MJy/sr. J is blue, H is green, and Ks is red. Outflows from young stars glow very red, while a normal reflection nebula in the upper-right is blue-green. Cloudshine, in contrast, shows here as a muted glow with green edges. Dark features around extended bright objects (such as the fan-shaped nebula in the upper-right) are the result of sky-subtraction.}
\end{figure}

\clearpage

\begin{figure}
\figurenum{2}
\plotone{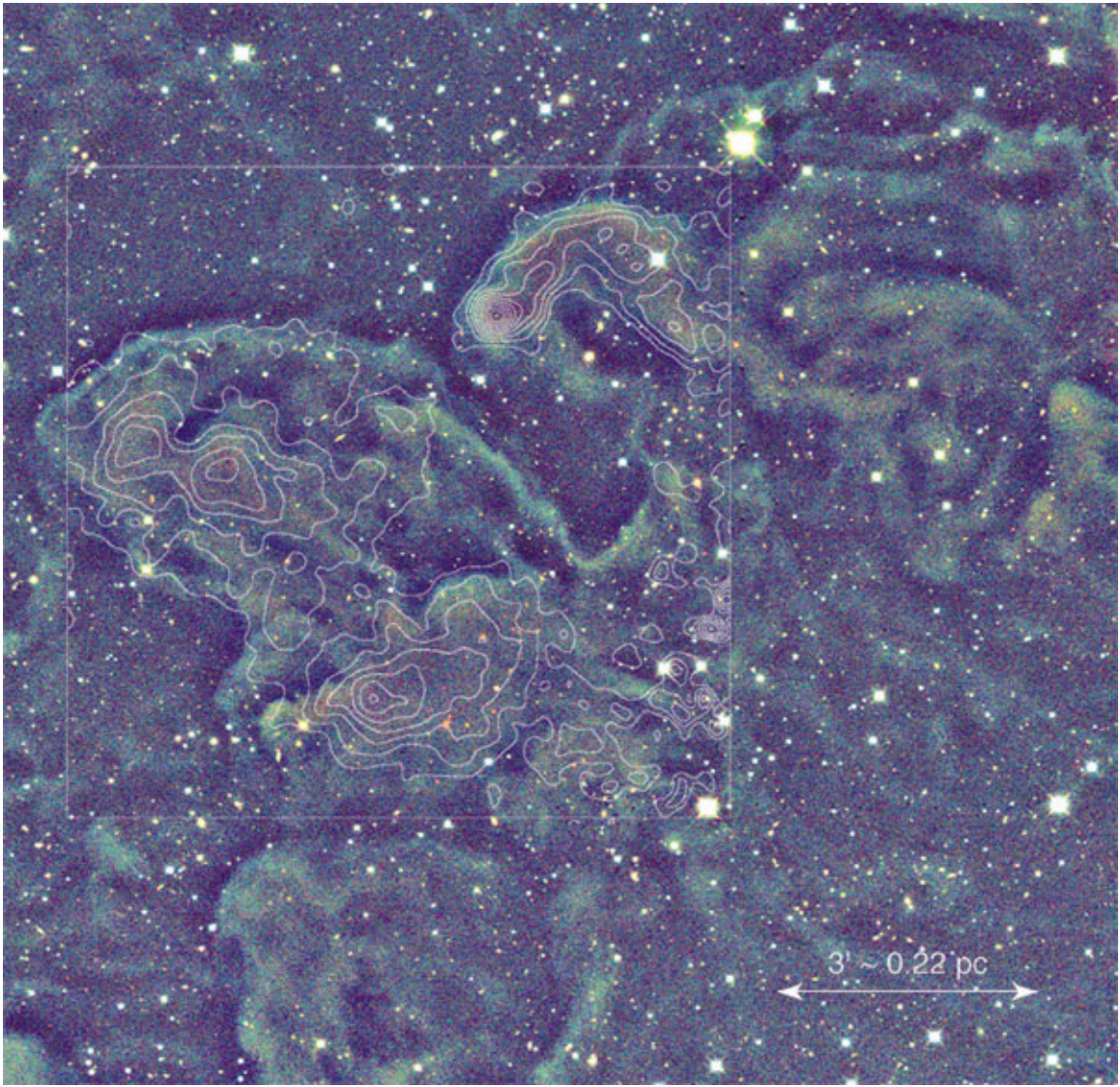}
\caption{L1451 in false color. Again, each component image has been scaled to the same flux scale in MJy/sr and J is blue, H is green, and Ks is red. A smaller map of 1.2-mm dust emission contours from COMPLETE (\citet{Tafalla2006}) has been overplotted. This shows that the color of cloudshine is a good way of seeing the density structure, as redder regions have high dust continuum flux, and the edges of cloudshine line up well with the edges of the dust emission.}
\end{figure}

\clearpage

\begin{figure}
\figurenum{3}
\plotone{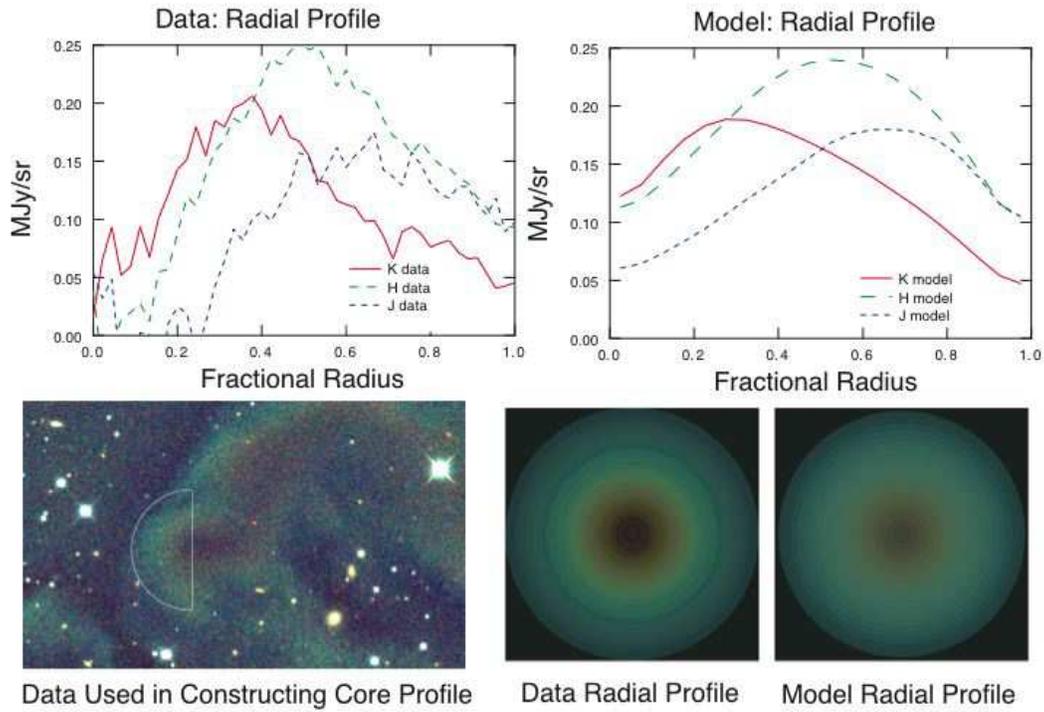}
\caption{A model of cloudshine in one core as reflected interstellar radiation. The lower-left figure shows the roughly circular feature we chose to model as a sphere. Due to the surrounding structure, only the left half of the circle was used to derive the density profile. The comparison between this density profile and our model fit is shown in two ways: above as radial flux profiles in individual bands and on the lower right as a synthetic color-composite image which allows overall comparison. Although the fit is roughly good, the central region of the core is much darker than predicted by the model. Some of this may be due to self-sky subtraction in the image and a non-spherical, non-isotropically illuminated core, and some may be due to a failure to adequately model the density structure at the center of the core.}
\end{figure}

\end{document}